\documentclass[article,referee]{aa}
\usepackage{subfigure}
\usepackage{natbib}
\usepackage{verbatim}
\usepackage{graphicx}
\usepackage{amssymb}  
\usepackage{amsmath}
\usepackage{auto-pst-pdf}
\usepackage[english]{babel}

\newcommand{\aftr}{}

\begin{document}

\title{Magnetic shuffling of coronal downdrafts}
 
\author{A. Petralia\inst{1,2} \and F. Reale\inst{1,2} \and S.Orlando\inst{2}}

\institute{Dipartimento di Fisica \& Chimica, Universit\`a di Palermo, Piazza del Parlamento 1, 90134 Palermo, Italy
\and INAF-Osservatorio Astronomico di Palermo, Piazza del Parlamento 1, 90134 Palermo, Italy}

\date{Rec;Acc}

\abstract
{Channelled fragmented downflows are ubiquitous in magnetized atmospheres, and have been recently addressed from an observation after a solar eruption.}
{We study the possible back-effect of the magnetic field on the propagation of confined flows.} 
{We compare two 3D MHD simulations of dense supersonic plasma blobs downfalling along a coronal magnetic flux tube. In one, the blobs move strictly along the field lines; in the other, the initial velocity of the blobs is not perfectly aligned to the magnetic field and the field is weaker.}
{The aligned blobs remain compact while flowing along the tube, with the generated shocks. The misaligned blobs are disrupted and merged by the chaotic shuffling of the field lines, and structured into thinner filaments; Alfven wave fronts are generated together with shocks ahead of the dense moving front.}
{Downflowing plasma fragments can be chaotically and efficiently mixed if their motion is misaligned to field lines, with broad implications, e.g., disk accretion in protostars, coronal eruptions and rain.}


\maketitle

\section{Introduction}

The corona is the outer part of the solar atmosphere. It is highly structured by the magnetic field, but is also highly dynamic: the flows are generated by various mechanisms. For instance, compressions and rarefactions at the footpoints can trigger up- or down-flows inside the magnetic channels  (e.g., spicules, siphon flows).
Depending on the speed, the upflowing plasma can be ejected outside the solar atmosphere \citep{Chen2011,Webb2012} and/or it falls back onto the surface \citep[e.g., ][ and references therein]{Innetal2012}. Downfalling fragments after an eruption were used as a template for the accretion in young stars both in the high-$\beta$ and low-$\beta$ regimes \citep{Reaetal2013,Reaetal2014,Petetal2016}. The accreting cold material from the circumstellar disk flows along magnetic channels and impacts the stellar surface \citep{Uchetal1984,Bertout1988}. The structure and dynamics of the falling material is presumably influenced both by the strength and complexity of the  magnetic fields and by the flow inhomogeneity \citep{Matetal2013,Orletal2013,Coletal2016}.  
Downfalls can be also generated by thermal instability \citep{Parker1953,Field1965} in the so-called coronal rain. In this case, a strong heating at loop footpoints can lead to a high plasma density in loops. The high radiative losses exceed the heating and cause a catastrophic plasma cooling and condensation \citep{Anteral2012,Kleetal2014,Fanetal2015}.

When the plasma falls in a region where the magnetic field is strong, it can be channelled along flux tubes. After a spectacular solar eruption in June 7, 2011, large fragments were spread all over the solar surface \citep{vanDrieetal2014,Innetal2012,Reaetal2013,Reaetal2014}, and \cite{Petetal2016} studied some falling close to active regions and strongly interacting with the magnetic field. Magnetohydrodynamic (MHD) modelling showed that the shocks ahead of the downfalling fragments brighten the final segment of the magnetic channel. The model showed also that the plasma blobs are warped and further fragmented as soon as their interaction with the field becomes significant. It is clear that the plasma is conditioned by the field and the field by the plasma.

In the present work we investigate this interaction and how it can or cannot determine a significant disruption of the blobs. To this purpose we compare two similar MHD simulations, one showing, the other not showing this effect.

\section{MHD Modelling}

\label{sec:themodel}

\begin{figure}[!t]
\centering
\includegraphics[width=5.5cm, angle=-90]{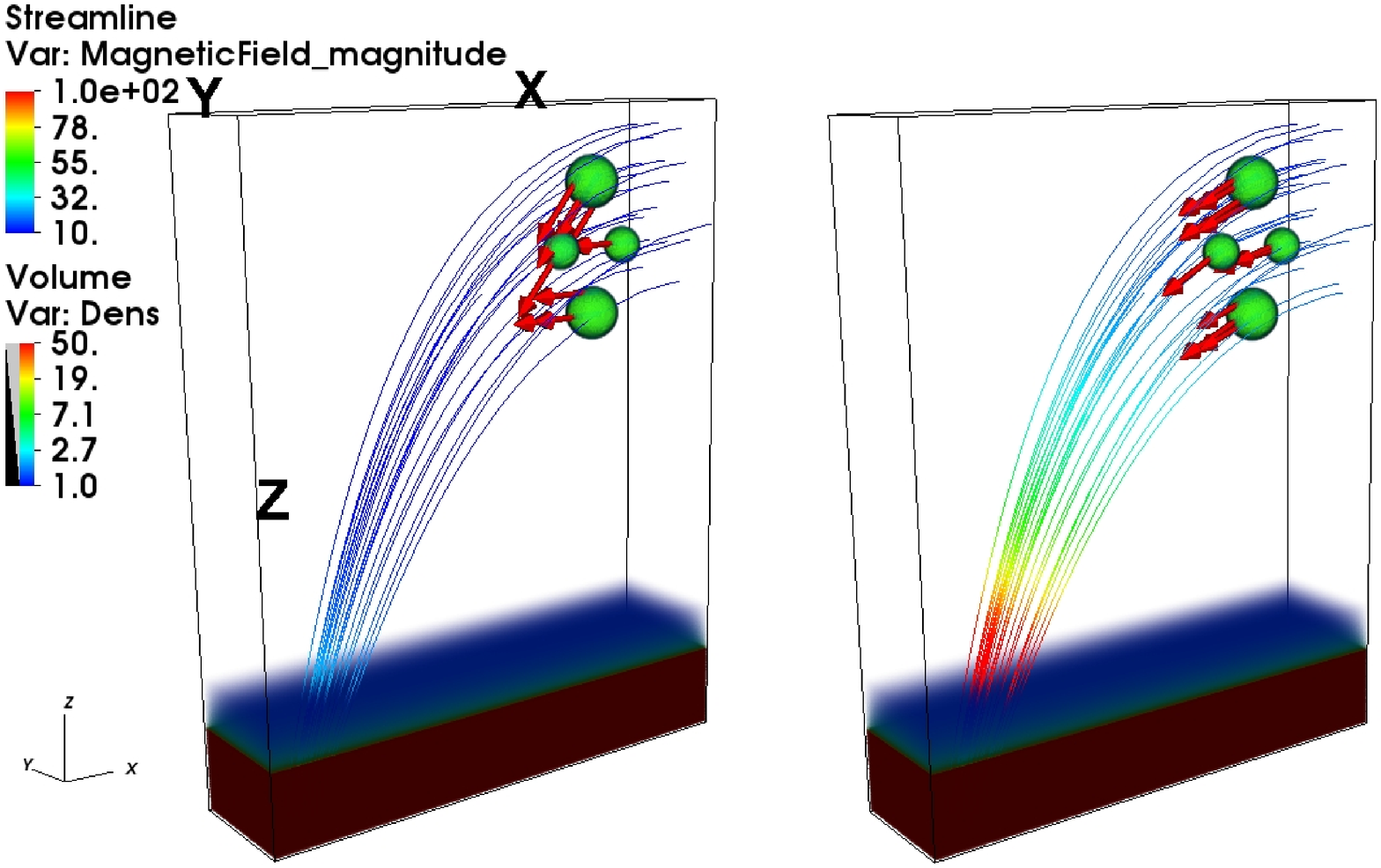}
\caption{Initial conditions of the two case simulations. \aftr{Volume} rendering of the density ($10^9$ cm$^{-3}$, logarithmic scale). Some magnetic field lines (G, \aftr{coloured by field intensity}) and blobs with initial velocity ({\it red arrows}) not aligned ({\it left}) and aligned ({\it right}) to the field lines are shown.}
\label{fig:inicond}
\end{figure}
 
As in \cite{Petetal2016}, we study the propagation of plasma blobs inside a magnetized corona through detailed MHD modeling. Our model solves the same MHD equations as described in \cite{Petetal2016}, including thermal conduction and radiative losses. The calculations are performed using the same MHD module available in {\it PLUTO}  \citep{MignBod07,MignZan12}, a modular, Godunov-type code for astrophysical plasmas. We use radiative losses from the CHIANTI code (Version 7) \citep{LanZan12}, assuming a density of $10^9$ cm$^{-3}$ and ionization equilibrium according to \citet{Dere09}. We assume \aftr{no losses and heating} in the chromosphere and inside the initial cold blobs (i.e., for $T \leq 10^4$ K). 

\begin{figure*}[!t]
\centering  
\includegraphics[width=5.85cm, angle=-90]{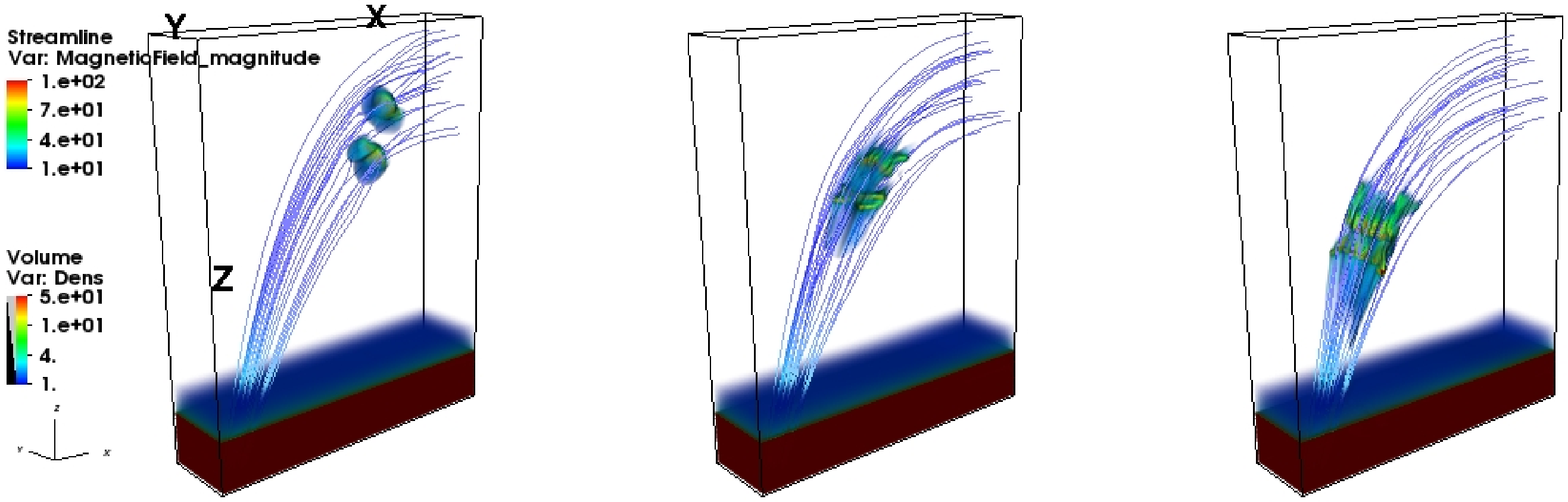}
\caption{Simulation of blobs not fully channelled by the magnetic field: \aftr{volume rendering of the} density at times t=$20$, $60$, $100$s as in Fig.\ref{fig:inicond}. \aftr{The temporal evolution is available in the online addition.}}
 \label{fig:misalign}
 \end{figure*}  
  
\begin{figure*}[!t]
\centering
\includegraphics[width=5.85cm, angle=-90]{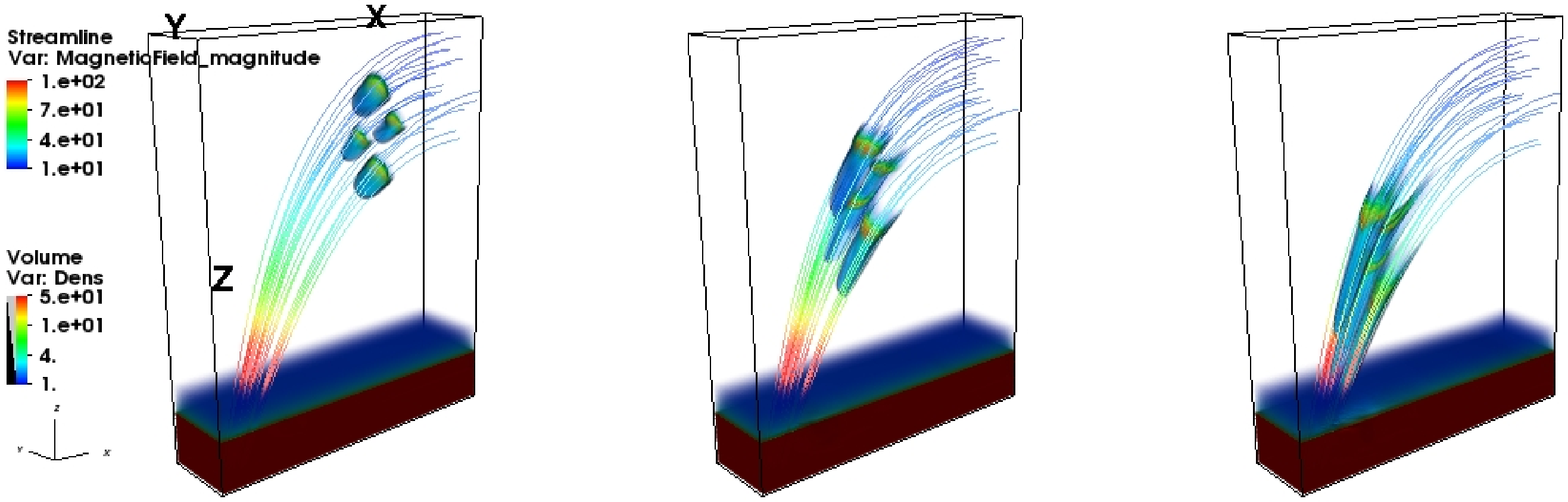}
\caption{Same as Fig.\ref{fig:misalign} but for the Simulation of blobs fully channelled by the field, \aftr{as well as for the temporal evolution.}}
 \label{fig:align}
 \end{figure*}
  
 	\begin{figure}[!t]
    \centering
    \includegraphics[width=5cm, angle= -90]{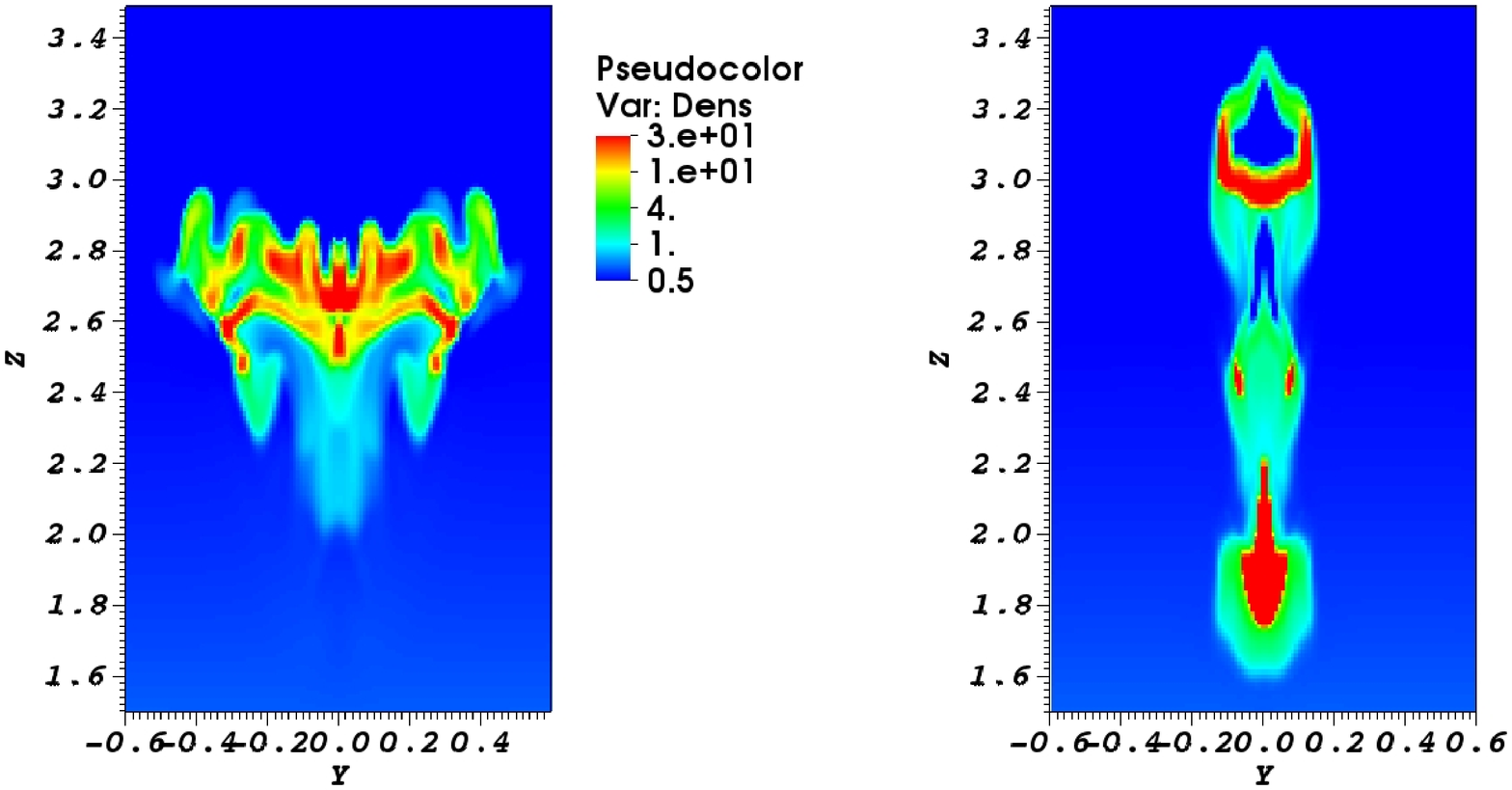}
\caption{Density ($10^9$ cm$^{-3}$, logarithmic scale) in a plane $YZ$ across the blobs for misaligned (left) at $t=100$ s \aftr{and $X=1.4\times10^9$cm}, and for aligned blob motion (right) at t=$90$s \aftr{and $X=1.5\times10^9$cm}.}
 \label{fig:comp_mod}
 \end{figure}

Our two simulations describe the evolution of four blobs moving across a magnetized  coronal atmosphere. {\it We compare a case in which the blobs are not fully channelled by the magnetic field to another in which they are.} We consider a typical coronal field configuration with closed arch-like lines anchored to the photosphere \citep{Reale2014}. This configuration has no special symmetry and a full 3D description is necessary. However, we can consider a symmetric magnetic field with respect to a plane perpendicular to the surface. Whatever the initial direction of the blobs, the field geometry and strength will prevent them from moving much across the field lines; therefore, the domain is not needed to be large in that direction, assumed to be the $Y$ direction. We approach the configuration of a loop-populated active region, still keeping it manageable and simple, with a combination of magnetic dipoles, such that the magnetic field is symmetric with respect to the side boundaries and is closed down in the chromosphere. The computational box is three-dimensional and cartesian $X$, $Y$, $Z$, and $4\times10^9$ cm, $1.2\times10^9$ cm, $6\times10^9$ cm long, respectively. The $Z$ direction is perpendicular to the solar surface. The mesh is uniformly spaced along the three directions with $512\times128\times512$ cells, and a cell size of $\sim80\times90\times120$ km, a good compromise between resolution in all directions (the domain is larger along $Z$) and computational times. The blobs are enough resolved (their diameter is 30-40 cells) and the initial atmosphere has been checked to be steady with this resolution. The ambient atmosphere is a stratified corona linked to a much denser chromosphere through a steep transition region.  The corona is a hydrostatic atmosphere \citep{RTV1978} that extends vertically for $10^{10}$ cm. The chromosphere is hydrostatic and isothermal at $10^4$ K and its density is $\sim 10^{16}$ cm$^{-3}$ at the bottom. The atmosphere is plane-parallel along $Z$. The coronal pressure ranges between $0.29$ dyn cm$^{-2}$ at the top of the transition region and $0.12$ dyn cm$^{-2}$ at $Z=10.5\times10^9$ cm. The density and temperature are, respectively, $\sim 2.2\times10^8$ cm$^{-3}$ and $\sim 2\times10^6$ K at $Z=10.5\times 10^9$ cm \citep{Reale2014}. The falling blobs and the atmosphere are very similar to one of the models in \cite{Petetal2016} ("Dense Model"), which are constrained from the observation and \aftr{therefore realistic}. Initially, the four blobs are at a height in the range $3.5 < Z < 4.5 \times10^9$ cm, and at a distance in a range $2.5 < X < 4 \times10^9$ cm from the left boundary side, close to the upper right corner. 

Fig.\ref{fig:inicond} shows our initial conditions. For simplicity's sake, we consider spherical blobs, with a radius between $1.4-2\times10^8$ cm, typical of those in the eruption of 7 June 2011, temperature $T = 10^4$ K and density $10^{10}$ cm$^{-3}$, \aftr{also typical of prominences \citep[e.g.][]{Labetal2010,Paretal2014}}. \aftr{We assume that the blobs are optically thick. We expect that the timescales of the radiative transfer from the blobs (not included in the model) to be much longer than that of the outside optically thinner plasma (included in the model), which in turn is much longer \citep[e.g.][]{Caretal1995,Reale2014} than the very short timescale of the dynamics (a couple of minutes only).} Their initial speed is $v = 300$ km/s. In one simulation, their motion is \aftr{cell-by-cell} aligned to the magnetic field. The field intensity is $\sim 170$ G at the top of the transition region and $\sim~15$ G at the initial height of the blobs. In the other simulation, the magnetic field is weaker here, i.e., $\sim 35$ G and $\sim 3$ G, respectively, and the speed of the blobs \aftr{is set uniform and} not totally aligned to the field but it lies in the $XZ$ plane: two have a horizontal initial direction, the other two have an inclination of $45^o$ downwards. 

Boundary conditions are reflective at the left end of the $X$ axis, the magnetic field is forced to be perpendicular to the boundary at the right end. For the other quantities, zero gradient has been set. Fixed conditions have been set at the lower end of the $Z$ axis, and zero gradient at the upper end, except for the magnetic field that is fixed. The same conditions are set at the far end of the $Y$ axis. The computational domain is symmetric to a plane in $Y=0$, so we simulate half domain and set reflective conditions at the lower end of the $Y$ axis.

We now describe the evolution of the flowing blobs, starting from those with an initial speed not aligned with the magnetic field, shown in Fig.~\ref{fig:misalign} and \aftr{the associated movie}. This case is similar to those illustrated in \cite{Petetal2016}, where the blobs were crossing a closed magnetic field while falling. The propagation along the field lines is presented for comparison and shows a striking qualitative difference from the other one, which is the main motivation for this work.  The initial speed of the blobs ($v=300$ km/s) is not far from a typical free-fall speed from large heights and larger than the local coronal sound speed ($c_s = \sqrt{\gamma p/\rho}~\sim~200$ km/s), so shocks are generated immediately. These are slow mode shocks that do not perturb the magnetic field and propagate along the magnetic field lines ahead of the blobs. \aftr{In spite of the initial temperature jump at the blob/corona transition region, the large difference in the heat capacity leads the conducted energy to be gradually radiated away, while the  blobs dynamics dominate all the evolution, as shown in previous work \citep{Petetal2016}}. However, the blobs themselves do not move parallel to the magnetic field, perturbing and warping it strongly in a few seconds. The ram pressure carried by the blobs is $p_{ram} = \rho v^2 \sim 20$ dyn cm$^{-2}$, much larger than the field pressure $B^2/8\pi \sim0.3$ dyn cm$^{-2}$; the magnetic tension gives the field enough stiffness to channel the blobs. The net effect is that the moving blobs produce a tailspin that travels along the field lines. Measuring the distance and time taken to arrive at the chromosphere, the speed of this perturbation is $\sim 700$ km/s, i.e., it is an Alfven wave that moves at an average  speed ($v_A = B/\sqrt{4\pi\rho}$) in a medium with density $7\times 10^8$ cm$^{-3}$ and magnetic field $\sim 10$ G, reasonable average conditions for the medium where the perturbation is propagating. \aftr{No MHD instabilities develop, the magnetic field is strong enough to suppress them (see Appendix \ref{sec:instab}).} While dragging the field lines, the misaligned and non-uniform motion of the blobs mixes them, and, as metal chords, they soon have a feedback on the blobs mixing them in turn. As a result, the blobs rapidly lose their initial shape and even their single identity. They first form two separate conglomerates in the initial 30~s, which travel along the tube. These are progressively squashed and elongated into a waterfall-like shape and in $\sim 2$ min they practically coalesce into a single blurred and filamented cloud, as shown in \aftr{Fig.~\ref{fig:misalign} and in the associated movie}. In the meantime, they still flow along the magnetic tube toward the chromosphere. The return shock from the chromosphere contributes to further disrupt and mix the downflowing cloud. At the end of the shuffling, the identity of the blob is completely lost, what remains is a highly inhomogeneous flow structured into filaments that move chaotically along the field lines until they hit the surface in $\sim200$~s. 

We have checked that we obtain a similar evolution both for blobs with diverging velocities and for a single blob with an initial speed not aligned to the magnetic field lines, i.e. the blobs are shuffled by the field and are disrupted.

Fig.~\ref{fig:align} and \aftr{the associated movie} show the propagation of blobs with a motion initially strictly aligned to the magnetic field lines. The velocity and the atmosphere conditions are equal to the previous case, so the generation and propagation of the slow mode shocks are the same: once they are generated, they propagate along the magnetic field lines. 
In this case,  the magnetic field intensity is five times greater than in the previous case, thus the magnetic field efficiently channels the blobs, and it is not perturbed significantly. The blobs simply flow along the magnetic field lines as the slow mode shocks do. No magnetic perturbation mixes the blobs, they remain compact during the motion. Their shape varies only because the magnetic channel changes its cross-section and direction along the propagation. The blobs do not merge and therefore do not lose their identity during the propagation, as clearly shown in Fig.~\ref{fig:align}. Fig.~\ref{fig:comp_mod} emphasises the difference between the evolution of the misaligned and aligned motions. It shows cross-sections of the density (Figs.\ref{fig:misalign} and \ref{fig:align}) in vertical $YZ$ planes. The images are taken at slightly different times, i.e., $t=100$ s and $90$ s, respectively, when the blobs are approximately in the same $Z$ range (the velocity component along the field lines is slightly different in the two cases). The figure shows very clearly how different is the evolution: a single but structured cloud versus three distant and separate blobs.

\section{Discussion and Conclusions}

In this work we study how different can be the propagation of fast plasma fragments flowing parallel to a coronal magnetic field from others flowing with a tilted direction, through detailed 3D MHD modeling. Here we use the same model as in \cite{Petetal2016} to describe the propagation of dense and cold blobs of plasma moving in a magnetized solar atmosphere (including both the chromosphere and the corona). The model includes the effect of the gravity, \aftr{optically thin} radiative losses, thermal conduction along the field lines, and magnetic induction. The magneto-hydrodynamic equations are solved numerically (PLUTO code) in 3D Cartesian geometry. 
We compare two similar simulations of blobs flowing inside a magnetic field anchored in the solar surface. In one, their motion is fully channelled by the magnetic field, in the other it is only partially, because of the initial direction of the motion and of the strength of the field. \aftr{We change the blobs initial conditions as less as possible from one case to the other. The compromise has been to use the same initial velocity, which the blobs may acquire when they flow inside very large arches, such as in huge prominences.}
The evolution that we find is strikingly different. In the fully aligned case, the blobs and the slow mode shocks flow along the field lines and do not perturb the intense magnetic field. The blobs remain compact and move inside independent magnetic channels. 
In the misaligned case, the shuffling of the field lines driven by the blobs has a feedback on the blobs themselves and mixes them. At the same time, the conglomeration is structured into thinner filaments. In this case ,it is impossible to establish what was the native channel or their initial shape, they lose completely their identity. Misaligned propagation is also an efficient way to excite fast Alfven wave fronts, which travel ahead of the cloud, in addition to shocks.

In summary, this work highlights the possible back-effect of the confining magnetic field on the propagation of fragmented flows inside it. If they are perfectly channelled, plasma fragments keep their identity as single blobs with no mixing, and the magnetic is left unchanged as well. If there is some misalignment, the magnetic field can react with a shuffling of the field lines that mixes and merges the fragments, thus changing completely the plasma configuration. This represents a very effective mechanism of plasma mixing in the presence of a magnetic field, different from standard shear-like instabilities. The field lines can be effectively shuffled by irregular plasma motion and its feedback to the plasma is naturally chaotic. One may wonder which is the most usual situation, whether aligned or misaligned fragment motion. We expect that if the plasma is confined since the beginning and the magnetic field does not change much along the track, the motion should be mostly aligned to the field \aftr{and even more if blobs' velocity is lower, as in the coronal rain, in which similar blobs fall by gravity and reach loop footpoints with a velocity of about 60 km/s \citep{Fanetal2013,Fanetal2015,Mosetal2015}. All our evolution occurs on timescales about two orders of magnitudes shorter than in these other studies, so our results might apply only to the very final stages of their modeling.}
On the other hand, downfalling from large distances through a significantly changing magnetic field might result into misaligned fragment motion. Such kind of situation may occur in the accretion onto young protostars from circumstellar disks, both at the flow origin (disk) and close to the flow impact, where the magnetic field of the star might become very complex. This process might therefore lead to further mixing of downflows and to increase their fine substructuring. 
\aftr{For this exploratory work, the initial conditions of our simulations differ both in the speed alignment and in the strength of the ambient magnetic field. Our aim here is just to show that the moving blobs can have two different destinies, but we do not explore the conditions to switch between the two in detail. This exploration is postponed to more extended work.}
Although this work mainly addresses downfall motions, it might be more general and, in particular, involve also the case of upflows, to be addressed in future work.

\acknowledgements{AP, FR, and SO acknowledge support from the Italian \emph{Ministero dell'Universit\`a e Ricerca}. PLUTO is developed at the Turin Astronomical Observatory in collaboration with the Department of Physics of the Turin University. We acknowledge the HPC facilities SCAN, of the INAF - Osservatorio Astronomico di Palermo, the CINECA Award HP10B59JKR for the availability of high performance computing resources and support.} 

\bibliographystyle{aa}
\bibliography{4blobs}

\newpage
\begin{appendix}

\section{MHD shear instabilities}
\label{sec:instab}

\subsection{Kelvin-Helmoltz instability}

When a heavier fluid (blobs) in motion is sustained against a lighter fluid (the corona) by the magnetic field, Kelvin-Helmoltz instabilities can arise. They are suppressed if the magnetic field is strong enough to satisfy the condition \citep[][and references therein]{Priest2014}
\begin{equation}
\frac{B_{-}^2+B_{+}^2}{4\pi \rho_{-}\rho_{+}} \left( \rho_{-}+\rho_{+} \right) \geq \left( U_{-}-U_{+} \right)^2
\label{eq:mhd_kh}
\end{equation}
\\
\noindent
where subscripts $-$ and $+$ denote the variables inside and outside the blobs, respectively, $B$ is the magnetic field, $\rho$ is the mass density, and $U$ is the velocity. 

Considering that the blob density ($n_b\approx10^{10}$cm$^{-3}$) is much higher than the coronal ambient density ($n_c\approx3\times10^{8}$~cm$^{-3}$), the magnetic field intensity is $\sim3$~G, for the misaligned blobs, and does not change much at the blob/corona interface, the ambient medium is static, Equation~\ref{eq:mhd_kh} can be simplified to 
\begin{equation}
\frac{2B^2}{4\pi n_b \mu m_H} \geq U^2  
\label{eq:mhd_kh1}
\end{equation}
\\
\noindent
where $\mu m_H$ is the mean atomic mass. We obtain  $3\times10^{15} > 10^{15}$ and an even larger difference for the aligned blobs where the magnetic field is 25 times more intense. Therefore, in our simulations Kelvin-Helmoltz instabilities are efficiently suppressed by the magnetic field. 

\subsection{Rayleigh-Taylor instability}

The high ratio between the density of the blobs and the ambient corona could make the separation blob/corona layer subject to the Rayleigh-Taylor instability. The wave vector of such perturbations is smaller a critical threshold given by  \citep[][and references therein]{Priest2014}

\begin{equation}
k<k_c=\frac{4\pi \left( \rho_{+}-\rho_{-} \right) g_{\sun}}{2B^2}
\label{eq:mhd_rt}
\end{equation}
\\
\noindent
with subscripts as in Eq.\ref{eq:mhd_kh}. This critical value leads to a lower limit for the characteristic length of the perturbation, that we estimated to be $L_c=2\pi/k_c > 10^{11}$~cm, which is much larger than the size of the blobs.

\end{appendix}

\end{document}